# Reduction of the effective thermal conductivity by circulation of the quasi-ballistic heat-flux

Ashok T. Ramu[1†], Carl D. Meinhart[2] and John E. Bowers[1]

1. Department of Electrical and Computer Engineering, University of California Santa Barbara, Santa Barbara, CA-93106, USA
2. Department of Mechanical Engineering, University of California Santa Barbara, Santa Barbara, CA-93106, USA

[†] Corresponding author: ashok.ramu@gmail.com

*Abstract: For a thermal conductivity that is spatially uniform and independent of temperature, Fourier's law of heat transfer predicts a curl-free heat-flux. In the quasi-ballistic phonon transport regime, where the Fourier law breaks down, it has been proven starting from the Boltzmann transport equation that the constitutive relation for the heat-flux contains a solenoidal (circulatory) term with zero divergence and non-zero curl. In this paper we show how to reduce the effective thermal conductivity of a material with dimensions on the same scale as the phonon mean-free path, by exploiting the solenoidal term in the constitutive relation.*

1. Introduction

Recent findings have established that the Fourier law of heat transfer breaks down at the micron scale in silicon and other crystalline materials, instead of at length-scales below a few tens of nanometers as predicted by kinetic theory arguments [1-3]. This is because the mean-free path spectrum of silicon is very broad, with phonons of mean-free paths a micron or larger contributing more than half of the net thermal conductivity [4].

Many efforts aimed at increasing the thermoelectric ('*ZT*') figure of merit of a material focus on reducing the thermal conductivity. Phonons of long wave-length and mean-free path are targeted by micro-structuring, for example through manipulation of grain-boundaries [5], while phonons of small wave-length and mean-free path are blocked by alloying [6] or by co-deposition of nanoparticles during growth [7-8]. The concomitant reduction of electron mobility is a concern with these approaches, and it is desirable for electron transport purposes to reduce the impact of long-wavelength phonons on the thermal conductivity while maintaining the crystallinity of the material.

In this paper, we propose to reduce the effective thermal conductivity by manipulating the patterns of quasi-ballistic heat flow. Our hypothesis is that when the heat-flux that enters the circulatory quasi-ballistic modes at the hot end, it does not equilibrate efficiently with the high-frequency, high heat-capacity phonon population at the cold end, and is therefore lost to the heat transfer process. Thus by forcing heat to circulate, we can reduce the effective thermal conductivity. We prove this hypothesis by solving the constitutive relation for the heat-flux simultaneously with energy conservation in two geometries, one where the lines of heat-flux are expected to travel almost directly from the hot end to the cold, and the other where they are expected to curve around an obstruction like a cavity.

2. The Theory

The theoretical development begins with the assumption of a simplified model of heat transfer, namely the two-channel model [9] wherein the phonon spectrum is divided into two parts- one a high-heat-capacity, high-frequency (HF) part that is in quasi-thermal equilibrium with a well-defined local temperature, and the other, a low-frequency (LF), low-heat-capacity part that is farther out of equilibrium. The LF modes do not interact with each other due to the small phase-space for such scattering [10], but can exchange energy with the HF modes. The LF modes are assumed to have a constant mean-free path. Upon iterating the Boltzmann transport equation for the LF modes, and inserting a tesseral harmonic expansion of the distribution function [11], we arrive after some manipulations at the following equation for the LF mode heat-flux [12]:

$$\boldsymbol{q}^{LF} = -\kappa^{LF}\nabla T + \frac{3}{5}(\Lambda^{LF})^2 \nabla(\nabla \cdot \boldsymbol{q}^{LF}) - \frac{1}{5}(\Lambda^{LF})^2 \nabla \times (\nabla \times \boldsymbol{q}^{LF}) \qquad (1)$$

The interested reader is referred to [12] for the detailed derivation of this equation. Here $\kappa^{LF}$ is the kinetic theory value of the thermal conductivity of the LF modes, and $\Lambda^{LF}$ is the LF mode phonon mean-free path. The third term on the right is of great interest here, predicting as it does a circulating quasi-ballistic heat-flux. We note that this term has no counterpart in one spatial dimension: thus two or three dimensional heat-flux is prerequisite for observing the effects predicted in this work.

The high-frequency (HF) modes are assumed to follow the Fourier law with a thermal conductivity $\kappa^{HF}$:

$$\boldsymbol{q}^{HF} = -\kappa^{HF}\nabla T \qquad (2)$$

We close the equations using energy conservation in steady-state:

$$\nabla \cdot (\boldsymbol{q}^{LF} + \boldsymbol{q}^{HF}) = 0 \qquad (3)$$

We assume white, perfectly specular boundaries for the LF modes. If $\widehat{\boldsymbol{n}}$ is the outward normal at the boundary, by differentiability of the heat-flux, a physical quantity, and excluding surface-evanescent modes, we have that the tangential as well as normal heat-flux at the boundary is 0:

$$\boldsymbol{q}^{LF} \cdot \widehat{\boldsymbol{n}} = 0 \qquad (4)$$

$$\boldsymbol{q}^{LF} \times \widehat{\boldsymbol{n}} = \boldsymbol{0} \qquad (5)$$

Combining, we have $\boldsymbol{q}^{LF} = \boldsymbol{0}$ at the boundary. Please see the appendix for further explanation of boundary conditions (BCs) Eq. (4) and Eq. (5). For the HF modes, we prescribe a given incoming heat-flux over one part of the boundary that we designate the 'hot' side, and an equal, outgoing heat-flux over another part of the boundary designated the 'cold' side, with zero heat-flux over all other boundaries. We write this boundary condition as:

$$\boldsymbol{q}^{HF} \cdot \widehat{\boldsymbol{n}} = -\kappa^{HF}\nabla T \cdot \widehat{\boldsymbol{n}} = -Q_s \qquad (6)$$

$Q_s$ is the boundary heat-flux. Using equations (1), (2) and (3) with BCs (4), (5) and (6), we solve for the temperature *T* throughout. The temperature *T* at the hot/cold sides is inversely proportional to the effective thermal conductivity. Thus keeping the heat-flux constant, we compare the hot/cold side temperatures for different geometries with and without the solenoidal term – the third term on the right-hand side of Eq. (1).

3. Reformulation of the basal equations for numerical solution by the finite-element method

To the best of our knowledge the equations presented above cannot be solved analytically even for the relatively simple geometries considered here. Thus we need to reformulate the equations to render them suitable for finite-element method (FEM) simulations. We rewrite Eq. (1) as follows:

$$\boldsymbol{q}^{LF} = -\kappa^{LF}\nabla T + \frac{3}{5}(\Lambda^{LF})^2\nabla(\nabla \cdot \boldsymbol{q}^{LF}) - \beta\frac{1}{5}(\Lambda^{LF})^2\nabla \times (\nabla \times \boldsymbol{q}^{LF}) \qquad (7)$$

Here setting $\beta = 1$ recovers the original equation, while setting $\beta$ to a small value like $10^{-5}$ will help us compare the results in two cases, with and without the solenoidal term. We prefer not omitting the term entirely by setting $\beta = 0$, because doing so will force us to omit one of the boundary conditions as well, in order to avoid over-constraining the problem. Thus setting $\beta$ to precisely 0 will not enable a fair comparison between the two cases.

Rearranging Eq. (7),

$$-\beta\frac{1}{5}(\Lambda^{LF})^2\nabla^2\boldsymbol{q}^{LF} + \boldsymbol{q}^{LF} = -\kappa^{LF}\nabla T - \frac{3-\beta}{5}(\Lambda^{LF})^2\nabla(u^{HF}) \qquad (8)$$

According to Eqs. (4) and (5), the boundary condition for this inhomogeneous vector Helmholtz equation is $\boldsymbol{q}^{LF} = 0$. Here $u^{HF} \equiv \nabla \cdot \boldsymbol{q}^{HF}$ per definition, and Eq. (8) follows from energy conservation, Eq. (3), i.e. $u^{HF} = -\nabla \cdot \boldsymbol{q}^{LF}$. However, in order to smooth out the third order derivatives on the right side of Eq. (8), we treat $u^{HF}$ as a separate, auxiliary variable, to be solved for. Using a small numerical damping parameter $\alpha$=0.01 micron, we write:

$$-\alpha^2\nabla^2 u^{HF} + u^{HF} = -\kappa^{HF}\nabla^2 T \qquad (9)$$

Eq. (9) follows from $\nabla \cdot \boldsymbol{q}^{HF} = -\kappa^{HF}\nabla^2 T$ (see Eq. (2)). The boundary condition for this equation is

$$u^{HF} = -\frac{3}{5}\frac{\kappa^{HF}}{\kappa^{HF}+\kappa^{LF}}(\Lambda^{LF})\nabla^2(\nabla \cdot \boldsymbol{q}^{LF}) \qquad (10)$$

and may be obtained by taking the divergence of Eq. (7).

Finally, the temperature is obtained from:

$$-\kappa^{HF}\nabla^2 T = -\nabla \cdot \boldsymbol{q}^{LF} \qquad (11)$$

The boundary condition for this equation is the prescribed heat-flux:

$$-\kappa^{HF}\nabla T.\widehat{\boldsymbol{n}} = -Q_s \qquad (12)$$

Equations (8), (9) and (11) together with their corresponding boundary conditions are solved iteratively using the finite element method implementation COMSOL®.

## 4. Results and discussion

We consider an infinitely long hollow silicon cylinder with the HF mode thermal conductivity $\kappa^{HF}$=30 W/m-K, an LF mode thermal conductivity $\kappa^{LF}$=60 W/m-K (bulk thermal conductivity = 90 W/m-K) and LF mode mean-free path = 0.5 micron [13]. The prescribed heat-flux is taken to be 0 on the inner cylindrical surface, and equal on the outer cylindrical surface to

$$Q_s = Q_0 sin\theta \qquad (13)$$

where $Q_0=10^8$ W/m² and $\theta$ is the angular variable in cylindrical polar coordinates. Thus heat is fluxed into the half-cylinder $\theta = 0 \to \pi$ and flows out of the half-cylinder $\theta = \pi \to 2\pi$.

We fix the outer radius of the cylinder to 2 micron and vary the inner radius and the parameter $\beta$ controlling the circulation of heat-flux. First consider the case where inner radius is small, 0.1 micron. Fig. 1a and Fig. 1b show the temperature profile for $\beta = 10^{-5}$ and $\beta = 1$ respectively. The hot-side ($\theta = \frac{\pi}{2}$) temperature varies from 2.23 K to 2.94 K above ambient as $\beta$ varies over this range.

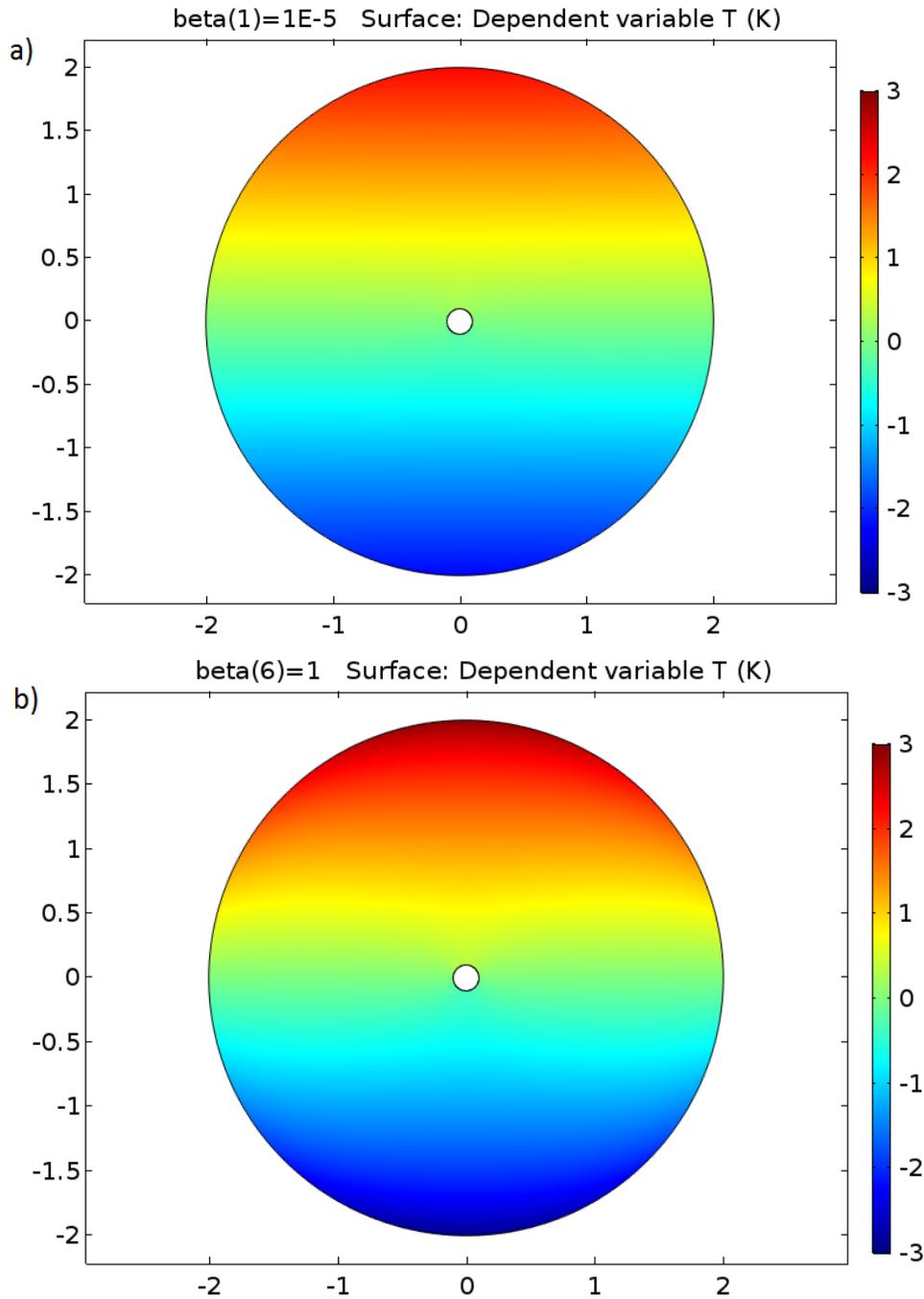

Fig. 1. Temperature profile in a hollow cylinder with inner radius 0.1 micron and outer radius 2 micron. (a) with circulation suppressed ($\beta = 10^{-5}$): maximum temperature=2.23 K above ambient, and (b) With circulation ($\beta = 1$): maximum temperature=2.94 K above ambient. All other parameters are as described at the beginning of this section.

In the geometry of Fig. 1, we might expect most of the heat transfer to take place directly from the top to the bottom, with the heat-flux mostly unidirectional; thus the differences between the non-circulating and circulating cases although present are modest – the effective thermal conductivity (as indicated by the hot-side temperature for a constant heat-flux) reduces by ~ 25%. This leads us to the idea of forcing the lines of heat-flux to curve around an obstruction like a cavity: one might expect the circulatory term to have a much larger effect in such a geometry. Simulations indeed confirm this expectation. Fig. 2 shows the temperature profile in a hollow cylinder with inner and outer diameter 1.5 microns and 2 microns. Noteworthy differences are seen between the non-circulatory and circulatory cases, with a 50% reduction in effective thermal conductivity due to the circulation of the heat-flux.

It may be argued that the results presented here are another manifestation of the well-known size effect, where the reduction of the characteristic length scale below the phonon mean-free path results in reduction of the effective thermal conductivity (see [4] for e.g.). That this is not the case may be deduced by noticing that even when $\beta = 0$, the term involving $\nabla(\nabla \cdot \boldsymbol{q}^{LF})$ remains in the constitutive relation, Eq. (7). We have established in earlier work [13][16] that this term when added to the Fourier law captures traditional size effects in transient grating and frequency-domain thermoreflectance experiments. Indeed, Fourier law simulation of the geometry of Fig. 2 yields a temperature variation of 7.9 K, which is less than the non-circulatory result, 8.09 K.

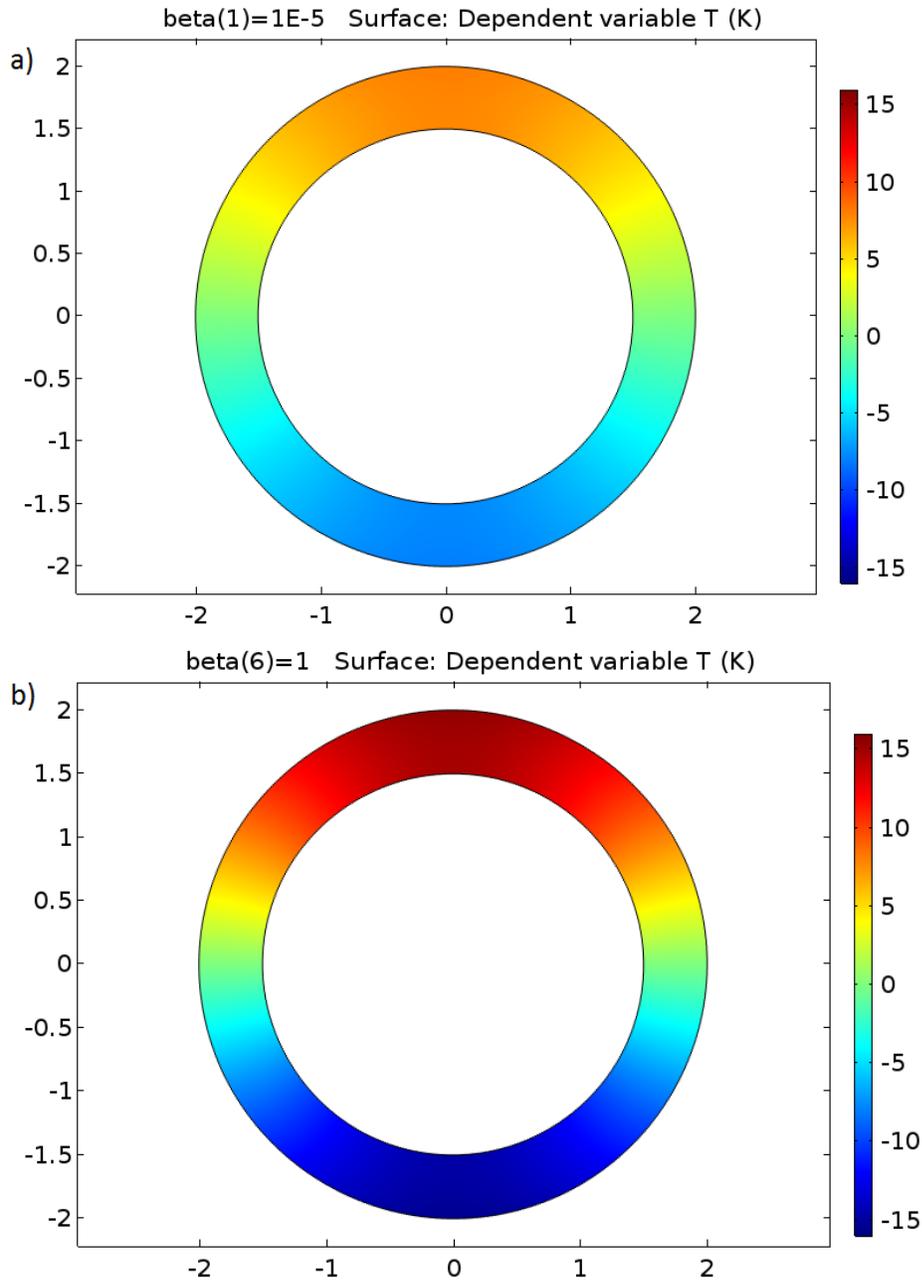

Fig. 2. Temperature profile in a hollow cylinder with inner radius 1.5 micron and outer radius 2 micron. (a) with circulation suppressed ($\beta = 10^{-5}$): maximum temperature=8.09 K above ambient, and (b) With circulation ($\beta = 1$): maximum temperature=15.5 K above ambient. All other parameters are as described at the beginning of this section.

Fig. 3 shows the plot of hot-side temperature vs. $\beta$: as $\beta$ reduces, the effective conductivity reaches a plateau, validating our numerical procedure of setting $\beta$ to a low value to investigate the case of non-circulatory heat-flux.

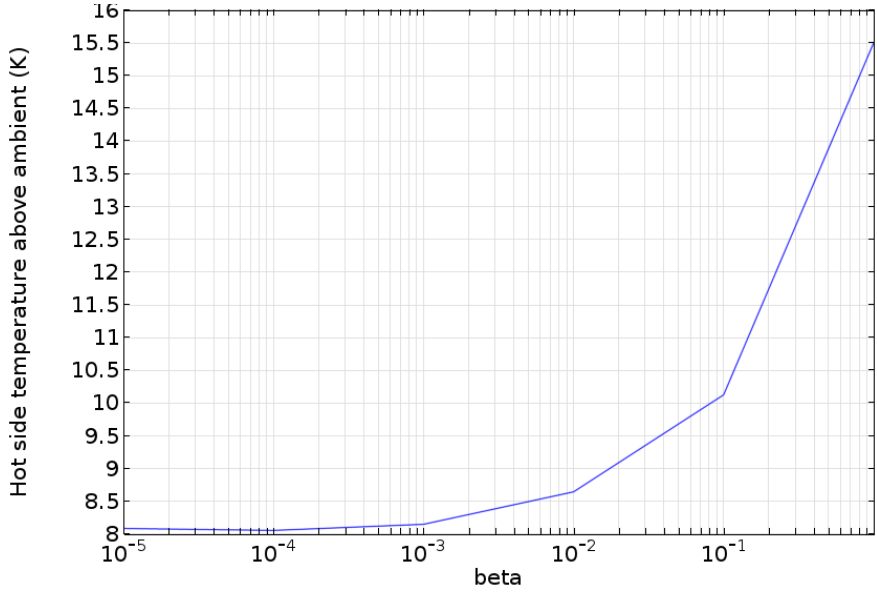

Fig. 3: Plot of hot side temperature vs. parameter $\beta$ showing plateauing for small $\beta$.

As mentioned in the introduction, we ascribe the reduction in thermal conductivity to the excitation of circulatory quasi-ballistic modes at the hot side that equilibrate poorly with the high-frequency reservoir at the cold side; thus forced circulation as in Fig. 2 implies a drastic reduction in the thermal conductivity. We further conjecture that the solenoidal quasi-ballistic heat-flux is responsible for the "necking" effect observed by Nomura *et al.* [14] and Tang *et al.* [15] to reduce the thermal conductivity of 2D phononic crystals.

## 5. Conclusions

In conclusion, we have shown that the circulatory (solenoidal) term in the heat-flux equations reduces the effective thermal conductivity in two or three dimensional structures. Numerical simulations show that the solenoidal term may be made more prominent by geometrically altering the pattern of heat-flux, resulting in further reduction of the thermal conductivity. These findings are expected to be of great interest to thermoelectric energy conversion, since they offer a route to reduction of the effective thermal conductivity without affecting the crystallinity or the electron transport properties of the material.

## Acknowledgments

We wish to thank Prof. Ali Shakouri (Purdue University) for helpful discussion. This work was funded by the National Science Foundation, USA under project number CMMI-1363207.

# Appendix 1: Proof of boundary conditions

We prove that the quasi-ballistic heat-flux vector satisfies the boundary condition $\boldsymbol{q}^{LF}=\mathbf{0}$ for specular, white boundaries.

Consider point P inside the computational domain close to the boundary, and point Q such that PQ is parallel to the surface tangent, as depicted in Fig. S1. Let the coordinates of P be $\boldsymbol{r}$ and those of Q be $\boldsymbol{r} + \boldsymbol{\delta r}.$

Suppose that the heat-flux through point $P_1$ is $\boldsymbol{q}^{LF}$ making an angle θ with the normal to the boundary (unit vector $\boldsymbol{n}$). We have that the heat-flux $\boldsymbol{q'}^{LF}$ through Q has the same magnitude as $\boldsymbol{q}^{LF}$ with direction as shown in Fig. S1, due to the specularity of the boundary.

The difference between the heat-flux at points Q and P is $q^{LF}(r + \delta r) - q^{LF}(r) = \nabla q^{LF} \cdot \delta r = q'^{LF} - q^{LF}$. But $q'^{LF} - q^{LF} = 2q\cos\theta n$ where $q$ is the magnitude of either $q'^{LF}$ or $q^{LF}$. Thus

$$\nabla q^{LF} \cdot \delta r = 2q\cos\theta n$$

Assuming continuity of each component of $q^{LF}$ we have that as $\delta r \to 0, q \to 0$ if $\theta \neq \pi/2$. Thus if $\theta \neq \pi/2$, the quasi-ballistic heat-flux $q^{LF}$ at the boundary has magnitude = 0, and therefore all components 0.

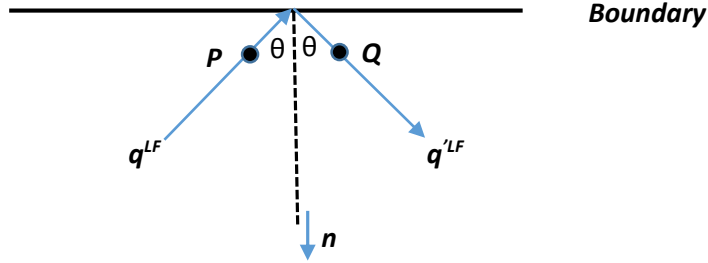

The remaining case, $\theta = \pi/2$ requires that heat-flux in the quasi-ballistic modes be parallel to the surface. Here we invoke an approximation of large anisotropy, which means that the heat-flux contribution at any frequency is dominated by phonon modes with wave-vectors parallel to the direction of the heat-flux. The farther out of equilibrium the phonons are, the better is this approximation. From the theory of elasticity, bulk phonons cannot exist parallel to a free boundary without violating the boundary condition that the normal stress vector be **0** [17]; thus the dominant contribution to the surface heat-flux is 0.

Thus in either case, if we exclude evanescent modes with imaginary wave-vector perpendicular to the boundary, we have that $q^{LF} = 0$.